\documentclass[twocolumn,preprintnumbers,amsmath,prc,amssymb]{revtex4}

\usepackage{graphicx}
\usepackage{dcolumn}
\usepackage{bm}
\usepackage{amsmath}
\usepackage{lipsum}
\usepackage{color}

\begin{document}

\title{Asteroseismology: radial oscillations of neutron stars with realistic equation of state}%

\author{V. Sagun$^{1,2}$, G. Panotopoulos$^{3}$, I. Lopes,$^{3}$}
\email[Emails: ]{violetta.sagun@uc.pt, grigorios.panotopoulos@tecnico.ulisboa.pt, ilidio.lopes@tecnico.ulisboa.pt}
\affiliation{$^1$ CFisUC, Department of Physics, University of Coimbra, 3004-516 Coimbra, Portugal}
\affiliation{$^2$ Bogolyubov Institute for Theoretical Physics, Metrologichna str. $14^{B}$, 03680, Kyiv, Ukraine}
\affiliation{$^3$ Centro de Astrof{\'i}sica e Gravita\c c\~ao-CENTRA, Departamento de F{\'i}sica, Instituto Superior T{\'e}cnico-IST, Universidade de Lisboa-UL, Av. Rovisco Pais 1, 1049-001, Lisboa, Portugal}

\begin{abstract}
We study radial oscillations of non-rotating neutron stars (NSs) in four-dimensional General Relativity. The interior of the NS was modelled within a recently proposed multicomponent realistic equation of state (EoS) with the induced surface tension (IST). In particular, we considered the IST EoS with two sets of model parameters, that both reproduce all the known properties of normal nuclear matter, give a high quality description of the proton flow constraint, hadron multiplicities created in nuclear-nuclear collisions, consistent with astrophysical observations and the observational data from the NS-NS merger.

We computed the 12 lowest radial oscillation modes, their frequencies and  corresponding eigenfunctions, as well as the large frequency separation for six selected fiducial NSs (with different radii and masses of 1.2, 1.5 and 1.9 solar masses) of the two distinct model sets. 
The calculated frequencies show  their continuous growth with an increase of the NS central baryon density.
Moreover, we found correlations between the behaviour of first eigenfunction calculated for the fundamental mode, the adiabatic index and the speed of sound profile, which could be used to probe the internal structure of NSs with the asteroseismology data.


\end{abstract}

\maketitle

\section{Introduction}
\label{Intro}

Compact objects, such as white dwarfs, neutron stars (NSs), hybrid or strange quark stars \citep{textbook,strangestars1,strangestars2}, are the final fate of stars, and they are characterized by ultra-high matter densities. NSs, in particular, are exciting objects as understanding their properties and their observed complex phenomena requires bringing together several different scientific disciplines and lines of research, such as nuclear physics, astrophysics and gravitational physics. These ultra-dense objects, thanks to their extreme conditions, which cannot be reached on earth-based experiments, constitute an excellent cosmic laboratory to study and constrain strongly interacting matter properties at high densities, phase transitions in it, non-conventional physics and alternative theories of gravity. 

It is well-known that the properties of compact objects, i.e. mass and radius, depend crucially on the equation of state (EoS), which unfortunately is poorly known. Presently, the main source of information about the properties of dense strongly interacting matter comes from the nucleus-nucleus (A+A) collisional programs which provide us with sufficiently accurate and detailed experimental data on the properties of nuclear and hadron matter at finite temperature. Using these data, however, it is highly non-trivial to formulate an EoS at vanishing temperature that corresponds to conditions inside the NSs. 

Another source of information comes from the merger of binary NSs. Thus, the LIGO/Virgo interferometers detection of gravitational waves emitted during the GW170817 NSs merger put constraints on the EoS at the super-high baryonic densities \citep{LIGO2017}. The precise timing of radio pulsars and X-ray observations of NSs in binaries led to progress in determination of masses and radii of compact stars \citep{Steiner2010, Steiner2013, Ozel2006}.

Combining all these pieces of information, an EoS with induced surface tension (IST) was recently proposed and tested \citep{LS2}. The IST EoS simultaneously reproduces existing heavy ion collision experimental data \citep{violetta2}, i.e AGS (Alternating Gradient Synchrotron, Brookhaven National Laboratory), SPS (Super Proton Synchrotron, CERN), RHIC (Relativistic Heavy Ion Collider, Brookhaven National Laboratory) and LHC (Large Hadron Collider, CERN) experiments, the nuclear matter properties \citep{violetta1}, the astrophysical and gravitational-wave observations, providing its applicability in the widest range of thermodynamic parameters \citep{LS2}. Moreover, it was successfully applied to the description of nuclear liquid-gas phase transition with the critical endpoint \citep{Ivanytskyi2017}. As was shown in \cite{LS2} the IST EoS reproduces NS properties and is fully consistent with all astrophysical data. 

On the other hand, asteroseismology is a widely used technique to probe the internal structure of stars which can be applied to NSs in order to study the thermodynamic properties inside the star. Studying the oscillations of stars and computing the frequency modes we can learn more about their composition and the EoS of the strongly interacting matter, since the precise values of the frequency modes are very sensitive to the underlying physics and internal structure of the star, see e.g. \cite{pulsating1, pulsating2, pulsating3, pulsating4, pulsating5, pulsating6, pulsating7, pulsating8, pulsating10, pulsating9} and references therein.

The oscillations in NSs can be excited by accretion, tidal forces in close eccentric binary system, starquakes caused by cracks in the crust, magnetic reconfiguration, during the supernova explosion or any other dynamical instabilities \citep{Franco2000,Tsang2012,Chirenti2017,Hinderer}. 

In the present work, we are interested in studying the radial oscillations of non-rotating NSs. For this purpose we selected six objects with different masses $M$ and radii $R$ (i.e., six fiducial stars of different radii and masses equal to 1.2, 1.5 and 1.9 $M_{\odot}$, which represent softer and stiffer EoSs). Performing a thorough analysis of the frequency of radial oscillation modes, that for a NS corresponds to radial acoustic modes, we were able to find a connection between the oscillation frequencies, and the thermodynamic properties (i.e., the EoS) of matter inside the NS.
Thus, searches for correlations between the oscillation modes and a strongly interacting matter EoS can help to probe an internal structure of the NSs and, especially, phase transitions in their interior, which is one of the primary targets of compact star physics.

In addition, this work is also very interesting due to its relevance for gravitational physics studies. Thus, despite the fact that radial oscillations of a spherical star do not emit gravitational waves, they can couple to the non-radial oscillations, amplifying them and producing gravitational radiation to a significant level \citep{pulsating5}. Moreover, it is well-known that  radial and non-radial oscillations share identical global properties. Therefore, by studying the global properties of the radial oscillations, we are also characterizing similar properties of the non-radial oscillations. This is, for instance, the case of a quantity known as the large separation -- {\it frequency difference} of oscillating modes with the same degree and consecutive radial order~\citep[e.g.,][]{LopesTC}. 

The next generation of the  gravitational wave detectors such as the Einstein Telescope or the Cosmic Explorer could detect such emission and provide information on the compact stars oscillations \citep{Chirenti2018}. Moreover, the launch of the eXTP \citep{Zand2019} and other following X-ray missions will also increase the expectations for the  detection of the NSs oscillations. When the detection of the radial oscillations of NSs will become possible, such connection can be used to constrain the EoS of compact stars with high precision.

\smallskip

Our work is organized as follows: in the next section we present a brief review of the hydrostatic equilibrium and structure equations, while the description of the EoS used in this study, i.e. the  IST EoS, is discussed in section 3. In section 4 we present the equations of radial oscillations in General Relatively, and  section 5 is devoted to the discussion of the numerical results obtained in this study. Section 6 is dedicated to the discussion of the excitation mechanisms of the oscillation modes and their detectability with the future gravitational wave detectors.
Finally, in section 7, we present the main conclusions of  our work. In this study, for convenience we use geometrical units ($\hbar=c=G=1$) and also adopt the mostly positive metric signature $(-+++)$.

\section{Hydrostatic equilibrium}
\label{HydEq}

We briefly review the structure equations for relativistic stars in 
General Relativity \citep[GR,][]{GR}. The starting point is
Einstein's field equations without a cosmological constant, which reads
\begin{equation}
G_{\mu \nu} = R_{\mu \nu}-\frac{1}{2} R g_{\mu \nu}  = 8 \pi T_{\mu \nu},
\end{equation}
where $g_{\mu \nu}$ is the metric tensor, $R_{\mu \nu}$ is the Ricci tensor, $G_{\mu \nu}$ is the Einstein tensor, and $R=g^{\mu \nu} R_{\mu \nu}$ is the Ricci scalar. The matter is assumed to be a perfect fluid with a stress-energy tensor given by
\begin{equation}
T_{\mu \nu} = P g_{\mu \nu} + (\rho + P) u_\mu u_\nu\equiv P g_{\mu \nu} + \zeta u_\mu u_\nu,
\end{equation}
$\rho$ is the energy density, $P$ is the pressure and $u_\mu$ is the four-velocity of the fluid. 
For convenience of notation, we have also introduced the function $\zeta (r)$ that is given by the expression:
$\zeta=\rho+P$.

\smallskip
As usual for non-rotating objects we seek static spherically symmetric solutions assuming for the metric the ansatz
\begin{equation}
ds^2 = -f(r) dt^2 + g(r) dr^2 + r^2 (d \theta^2 + sin^2 \theta d \phi^2),
\end{equation}
where $f(r)$ and $g(r)$ are two unknown metric functions, that can also be written as 
$f(r)=e^{\lambda_1(r)}$ and $g(r)=e^{\lambda_2(r)}=(1-2 m(r)/r)^{-1}$.
Accordingly, the solutions inside and outside a compact star are obtained from the match of the following equations:
\begin{itemize}
\item[$-$]  For the  interior of the star ($r < R$), it is convenient to work with the functions $\lambda_1(r)$ and $m(r)$,
 instead of the functions $f(r)$ and $g(r)$. Thereby, the Tolman-Oppenheimer-Volkoff equations \citep[TOV,][]{TOV1,TOV2} 
 for the interior solution of a relativistic star read 
\begin{eqnarray}
m'(r) & = & 4 \pi r^2 \rho(r),
\label{eq:mprime}
\\
P'(r) & = & - \zeta(r) \frac{m(r)+4 \pi P(r) r^3}{r (r-2 r m(r) )}, 
\label{eq:Pprime}
\\
\lambda_1'(r) & = & -2 \frac{P'(r)}{\zeta(r)},
\label{eq:nprime}
\end{eqnarray}
where the prime denotes differentiation with respect to r. Moreover,  we  assume a certain EoS relating $P$ with $\rho$,
to obtain a closed system of differential equations. In the present work we consider the IST EoS \citep{LS2} 
as will be described in the next section.
 
\item[$-$] 
For the exterior of the star ($r > R$), the matter energy momentum tensor vanishes, 
and one obtains the well-known Schwarzschild solution~\citep{SBH} that reads
\begin{equation}
f(r) = g(r)^{-1} = 1-\frac{2 M}{r}.
\end{equation}
\end{itemize}
The first two equations (\ref{eq:mprime} and \ref{eq:Pprime}) are to be integrated with the initial conditions $m(r=0)=0$ and $P(r=0)=P_c$, where $P_c$ is the central pressure. The radius of the star is determined requiring that the pressure vanishes at the surface, $P(R) = 0$, and the mass of the star is then given by $M=m(R)$.
Moreover, it is required that the two solutions  match  at the surface of the star.
Finally, the other metric function can be computed using the third equation (i.e., equation~\ref{eq:nprime}) together with the boundary condition $\lambda_1(R)=ln(1-2M/R)$.


\section{Induced surface tension equation of state (IST EoS)}
\label{EOS}

The computation of the  TOV equations in a closed form, as well as the calculation of the NS radial oscillation modes require a relation between pressure and energy density, which is given by the EoS. For this purpose we use the IST EoS first formulated for symmetric nuclear matter by \cite{violetta1}. Furthermore, this EoS was formulated for $\beta$-equilibrated electrically neutral nucleon-electron mixture and applied to the NS modelling \citep{LS1, LS2}. Here, we use the most recent and advanced version of the IST EoS, which also accounts for the nuclear asymmetry energy \citep{LS2}.

In the Grand Canonical Ensemble the IST EoS has the form of the system of two coupled equations for the pressure $p$ and the IST coefficient $\sigma$:
\begin{eqnarray}
\label{Eq1}
p&=&
\hspace*{-.3cm}\sum_{A=n,p,e}p^{id}(m_A, \nu^1_A)
-p_{int}(n^{id}_B)+p_{sym}(n^{id}_B,I^{id}) \,,\quad\\
\label{Eq2}
\sigma&=&\hspace*{-.2cm}\sum_{A=n,p}p^{id}(m_A, \nu_A^2) R_{nucl} \, .
\end{eqnarray}
Neutrons, protons and electrons (subscript indexes $``n"$, $``p"$, and $``e"$, respectively) with corresponding masses $m_A$ and chemical potentials $\mu_A$ ($A=n,~p,~e$) are physical degrees of freedom explicitly included in the IST EoS. 
As was shown by the fit of A+A collision experimental data with the multicomponent IST EoS ~\citep{2018NuPhA.970..133B}, neutrons and protons are supposed to have the same hard core radii $R_{n}=R_{p}=R_{nucl}$, which lies in the range from 0.3 to 0.5 fm. For simplicity, interactions of electrons are neglected, and they are treated as free particles with a zero hard core radius $R_e=0$.

The system of Eqs. (\ref{Eq1}-\ref{Eq2}) is written in terms of the zero temperature pressure $p^{id}$ of non-interacting Fermi particles with spin $\frac{1}{2}$ and quantum degeneracy $2$, as

\begin{eqnarray}
\label{Eq3}
p^{id}(m,\mu) \hspace*{-0.05cm} = \hspace*{-0.05cm}
\frac{\mu k(2\mu^2-5m^2)+3m^4\ln\frac{\mu+k}{m}}{24\pi^2}\theta(\mu-m),\quad
\end{eqnarray}
where $k=\sqrt{\mu^2-m^2}$ is the Fermi momentum of a particle with mass $m$ and chemical potential $\mu$, and  $\theta$ is the Heaviside function. 

Interaction between nucleons accounts via a short range repulsion of the hard core type controlled by their hard core radius $R_{nucl}$ and the mean-field type attraction. Such an attraction leads to a negative shift of the one particle energy levels or, equivalently, to a positive contribution $U$ to the effective chemical potential of each nucleon  $\nu_A^1$ and $\nu_A^2$ ($A=p,n$). These effective chemical potentials include the effects of the hard core repulsion through the nucleon eigenvolume ${\rm V}=\frac{4}{3}\pi R_{nucl}^3$ and surface $S=4\pi R_{nucl}^2$, whereas the mean field attraction and symmetry energy are accounted for through the density and nucleon asymmetry dependent potentials $U$ and $U_{sym}$, respectively~\citep{Rischke1988}. Thus,
\begin{eqnarray}
\label{Eq5}
\nu^1_A&=&\mu_A-p{\rm V}-\sigma S+U(n^{id}_B) \mp U_{sym}(n^{id}_B,I^{id}),\\
\label{Eq6}
\nu_A^2&=&\mu_A-p{\rm V}-\alpha\sigma S+U_0.
\end{eqnarray}
Requirement of thermodynamic consistency leads to appearance of the mean-field contribution $p_{int}$ to the total pressure. Note, that $p_{int}$ enters the expression for the pressure with sign ``-'', since it is caused by the nucleon attraction. These two quantities $U$  and $p_{int}$ (which are controlled by constant parameters $C_d^2$ and $\kappa$) 
are written explicitly in the following form
\begin{equation}
U(n^{id}_B)=C_d^2 ~(n^{id}_B)^\kappa,\quad p_{int}(n^{id}_B)=\int\limits_0^{n^{id}_B} l~
\frac{\partial U(l)}{\partial l}~dl,
\end{equation}
where $n^{id}_B$ is the density of the baryonic charge. More precise account for the nucleon attraction leads to an additional positive shift of the particle chemical potential for $U_0=const$, which, however, does not contribute to the pressure due to its constant value. 
The nuclear symmetry energy contribution in the IST EoS is also taken into account within the mean-field theory framework. 
However, it corresponds to the nucleon repulsion, the contribution to the total pressure $p_{sym}$ enters it with sign ``+''. 
Note, that the shifts of the nucleon and proton chemical potentials by modulus are equal to $U_{sym}$, but have the opposite signs,
``-'' and ``+''  for neutrons and protons, respectively. In comparison to the parameterization of $p_{sym}$ in Ref. \citep{LS2}, that gives 
nuclear asymmetry energy slope at nuclear saturation density $L\sim 113-115$ MeV being on the limit of its value constrained by experiments \citep{Symslope}, here we considered another parameterization, which, on the one hand, gives lower value of $L$, and, on the other hand, do not violates the thermodynamic consistency \citep{Rischke1988, Bugaev1989}. In terms of the nuclear asymmetry parameter $I=(n^{id}_{n}-n^{id}_{p})/n^{id}_B$ ($n^{id}_{n}$ and $n^{id}_{p}$ are the densities of the ideal gas of neutrons and protons, respectively) it is parametrized as
\begin{eqnarray}
U_{sym}(n^{id}_B,I^{id}) &=&\int\limits_0^{n^{id}_B,I^{id}}  \frac{\partial p_{sym}(l)}{\partial l} \frac{dl}{l}
, \\
p_{sym}(n^{id}_B,I^{id}) &=& \frac{A_{sym}(n^{id}_B I^{id})^2}{\left(1+(B_{sym}n^{id}_B I^{id})^2 \right)^2},
\end{eqnarray}
where $A_{sym}$ and $B_{sym}$ are constants.

\begin{table*}
\begin{tabular}{|c|l|l|l|l|l|l|l|l|l|l|}
& $R_{nucl}$ & $\kappa$  & $B_{sym}$& $A_{sym}$ &$C_{d}^{2} $ & $U_0$  & $K_0$  & $L$ \\
Set &  $(fm)$ & $-$ & $(fm^{3})$ & $ (MeV \cdot fm^{3})$& $(MeV \cdot fm^{3\kappa})$& (MeV) & (MeV) & (MeV) \\
\hline
\hline
A & 0.477 & 0.254  & 14.0 & 111.87 & 145.90 & 157.35 &  202.36 & 96.05 \\
B & 0.463 & 0.25   & 16.0 & 138.30 & 146.30 & 162.87  &  201.02 & 93.19  \\
\end{tabular}
\caption{\label{tab1}Two sets of parameters of the IST EoS which reproduce the nuclear matter properties, the flow constraint and satisfy all astrophysical constraints. Moreover, the two IST EoS sets have the following common parameters: $\alpha=1.245$ and $J= 30.0\;{\rm MeV}$. }
\label{tab:ISTEoS}
\end{table*} 

\begin{figure}
\centering
\includegraphics[scale=0.45]{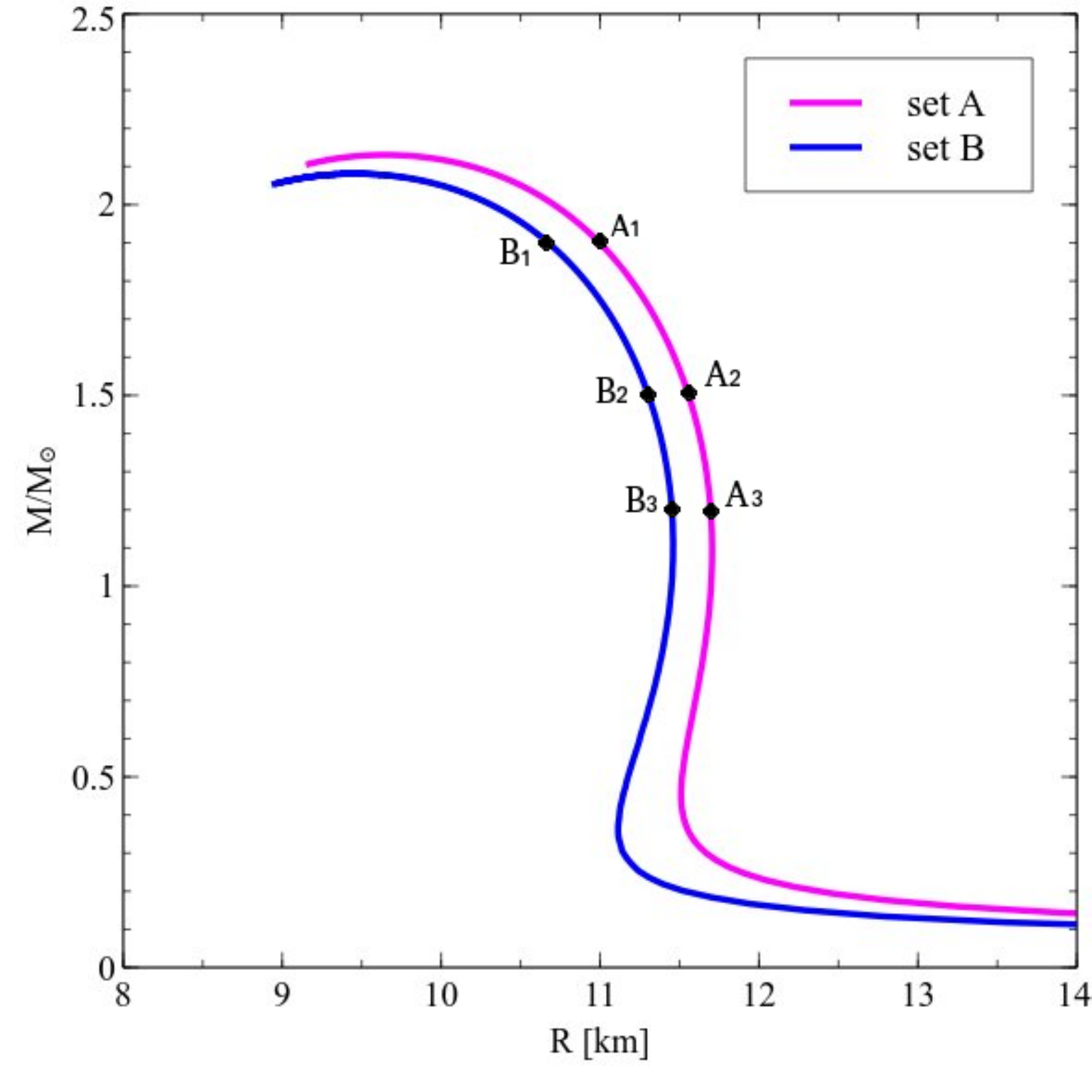}
\caption{The relation between gravitational mass M of NS and its radius R. 
The points correspond to the three fixed masses of the NS, i.e. $1.9 M_{\odot}$, $1.5 M_{\odot}$ and $1.2 M_{\odot}$.
}                                                                                                                                                                                                                   
\label{fig:massrad} 	
\end{figure}

\begin{figure}
\centering
\includegraphics[scale=0.26]{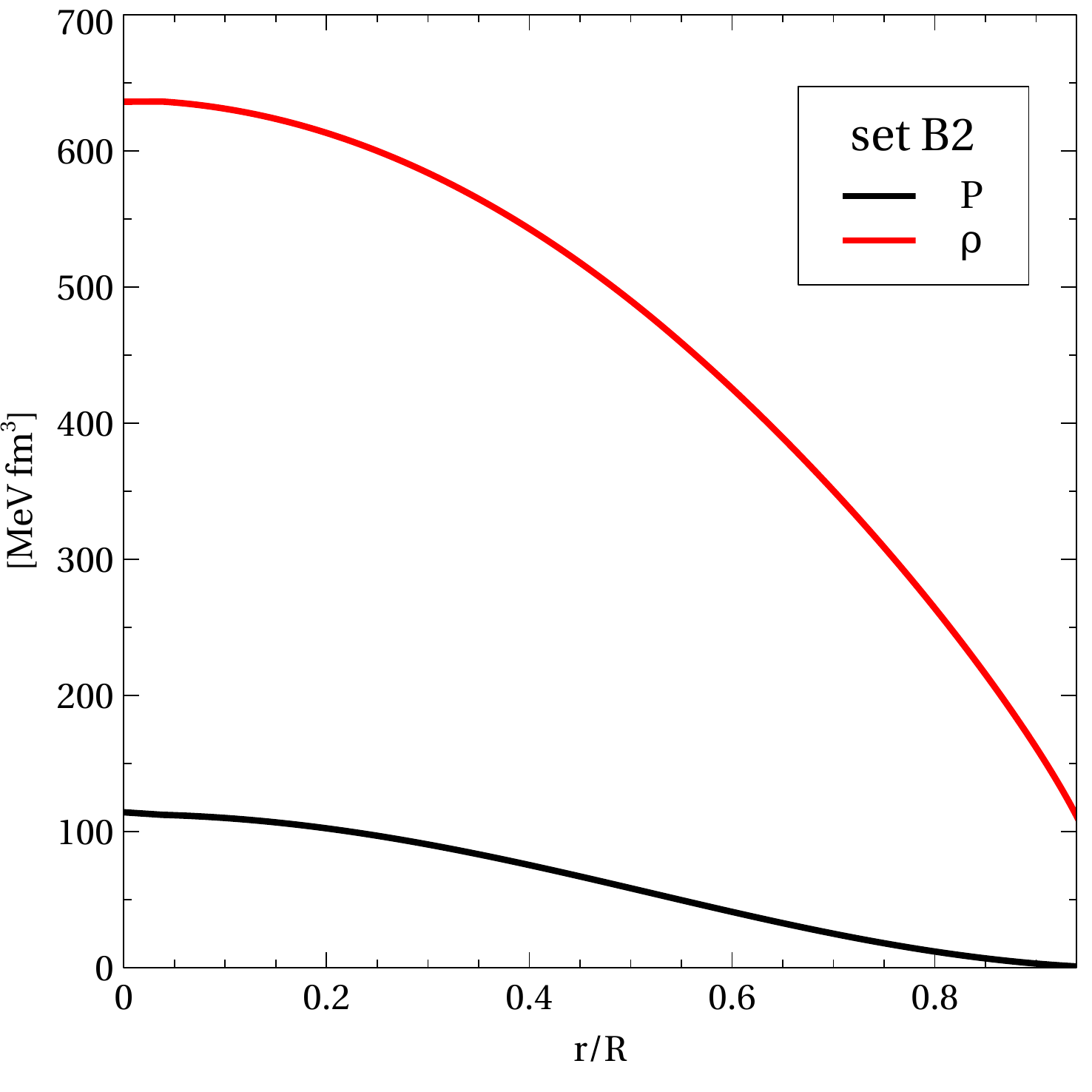} 
\includegraphics[scale=0.26]{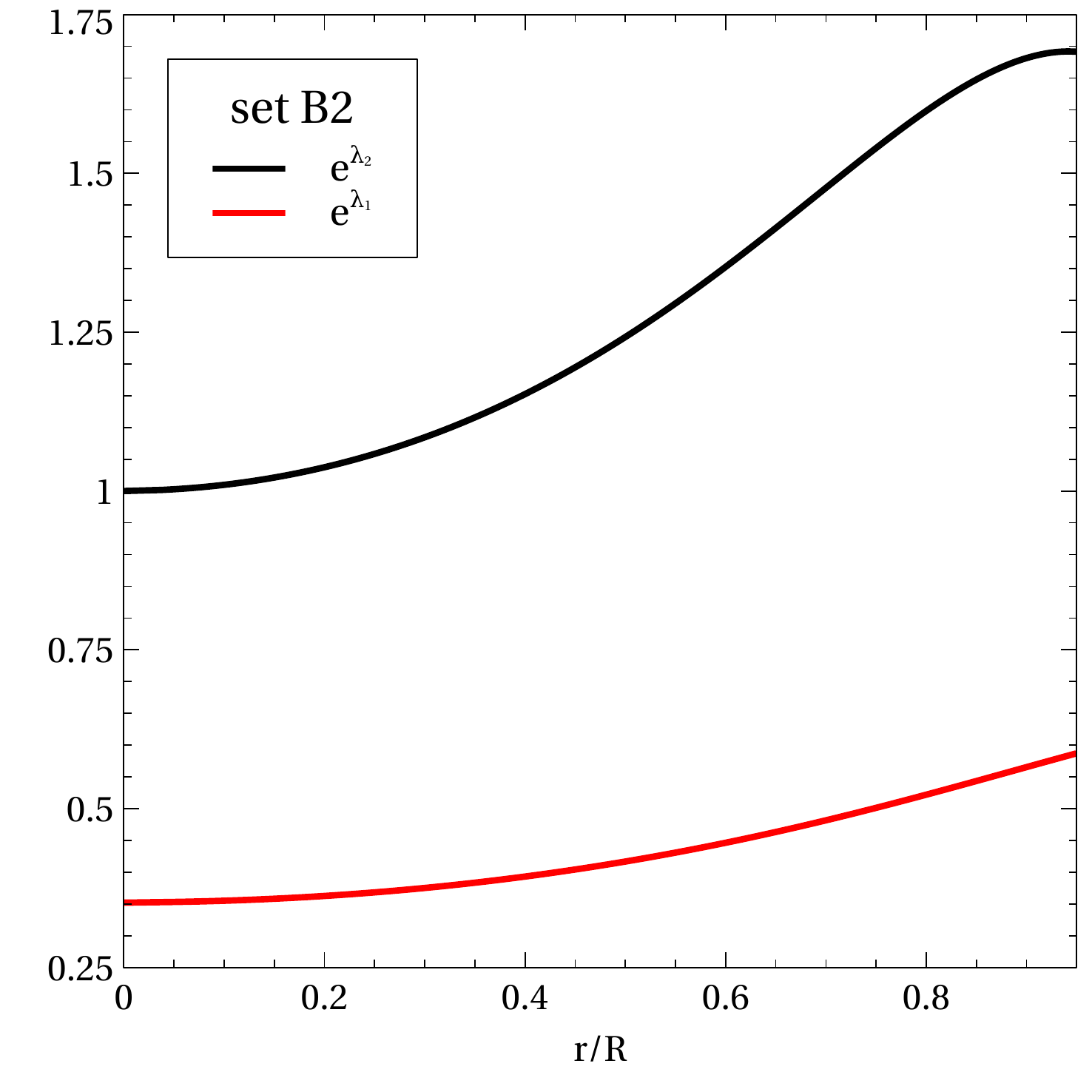} 
\caption{
{\bf Left panel:} Energy density $\rho$ (red curve) and pressure $P$ (black curve) as a function of the normalized radius $r/R$. {\bf Right panel:} Metric functions $e^{\lambda_1}$ (red curve) and $e^{\lambda_2}$ (black curve) as a function of $r/R$. Both panels correspond to $B_2$ star.}
\label{fig:interior} 	
\end{figure}

\begin{table}
\begin{center}
\begin{tabular}{l | l l l l l}
 & \multicolumn{5}{c}{{\sc Properties of six fiducial stars}} \\
Stars& $M (M_{\odot})$  & $R$ (km) & $\beta=M/R$ & $\omega_0$ (kHz) & $n_{B}^{c}$ ($fm^{-3}$)\\
\hline
\hline
 $A_1$  & 1.9 & 11.08 & 0.255  & 13.669 &  0.820 \\
 $A_2$  & 1.5 & 11.60 & 0.192 &  11.330 &  0.598\\
 $A_3$  & 1.2 & 11.73 & 0.152 &  9.965  &  0.480 \\
 $B_1$  & 1.9 & 10.67 & 0.265 &  14.453 &  0.866\\
 $B_2$  & 1.5 & 11.31 & 0.197 &  11.778 &  0.613\\
 $B_3$  & 1.2 & 11.45 & 0.156 &  10.333 &  0.491\\
\end{tabular}
\caption{Six fiducial stars are considered in this work from the two sets ($A$ and $B$) of the IST EoS.
The fundamental frequency $\omega_0$ is given by the expression $\omega_0=\sqrt{M/R^3}$. The last column represents central baryon densities for the selected stars.}
\label{tab:1set}
\end{center}
\end{table}

\begin{figure}
\centering
\includegraphics[scale=0.45]{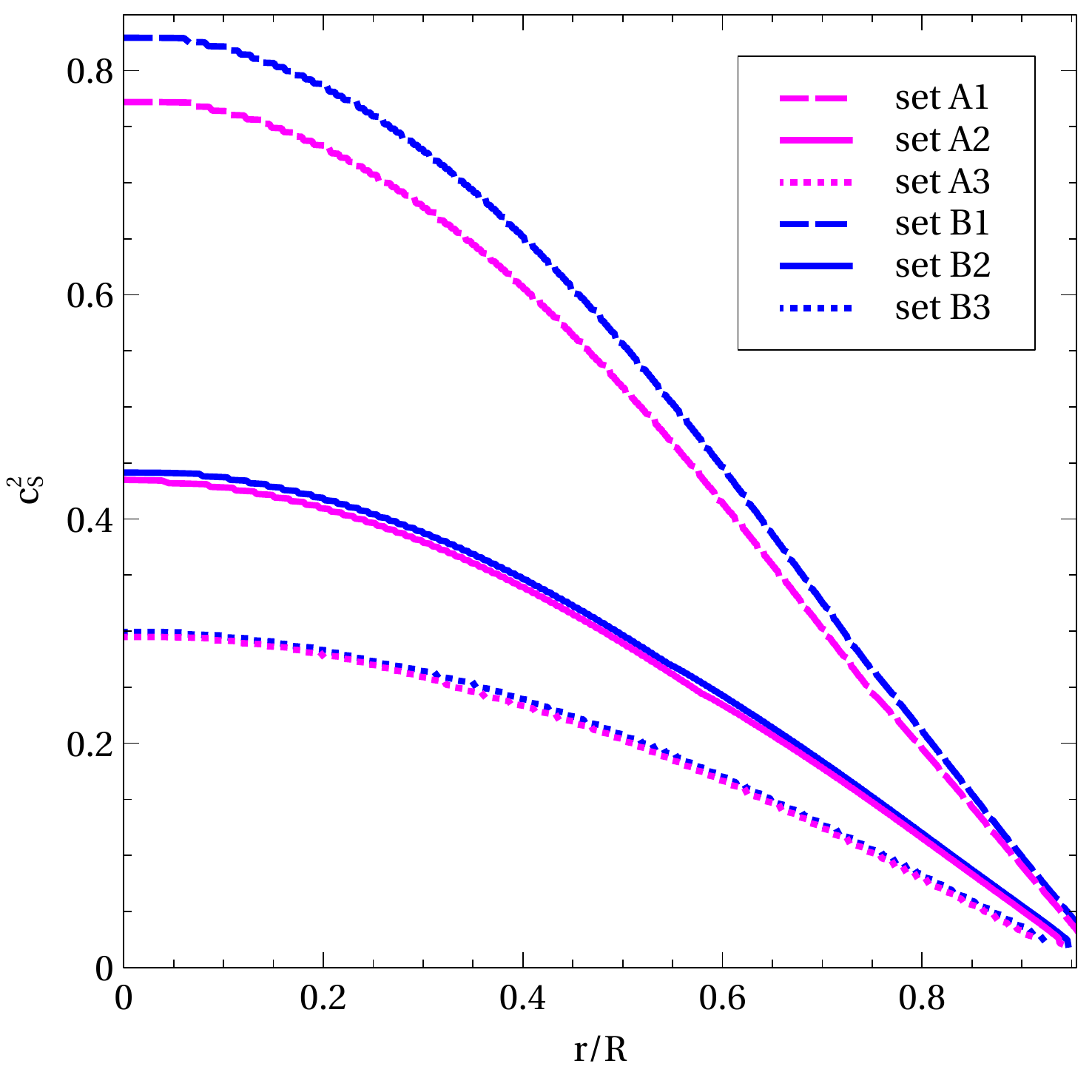} \\
\includegraphics[scale=0.45]{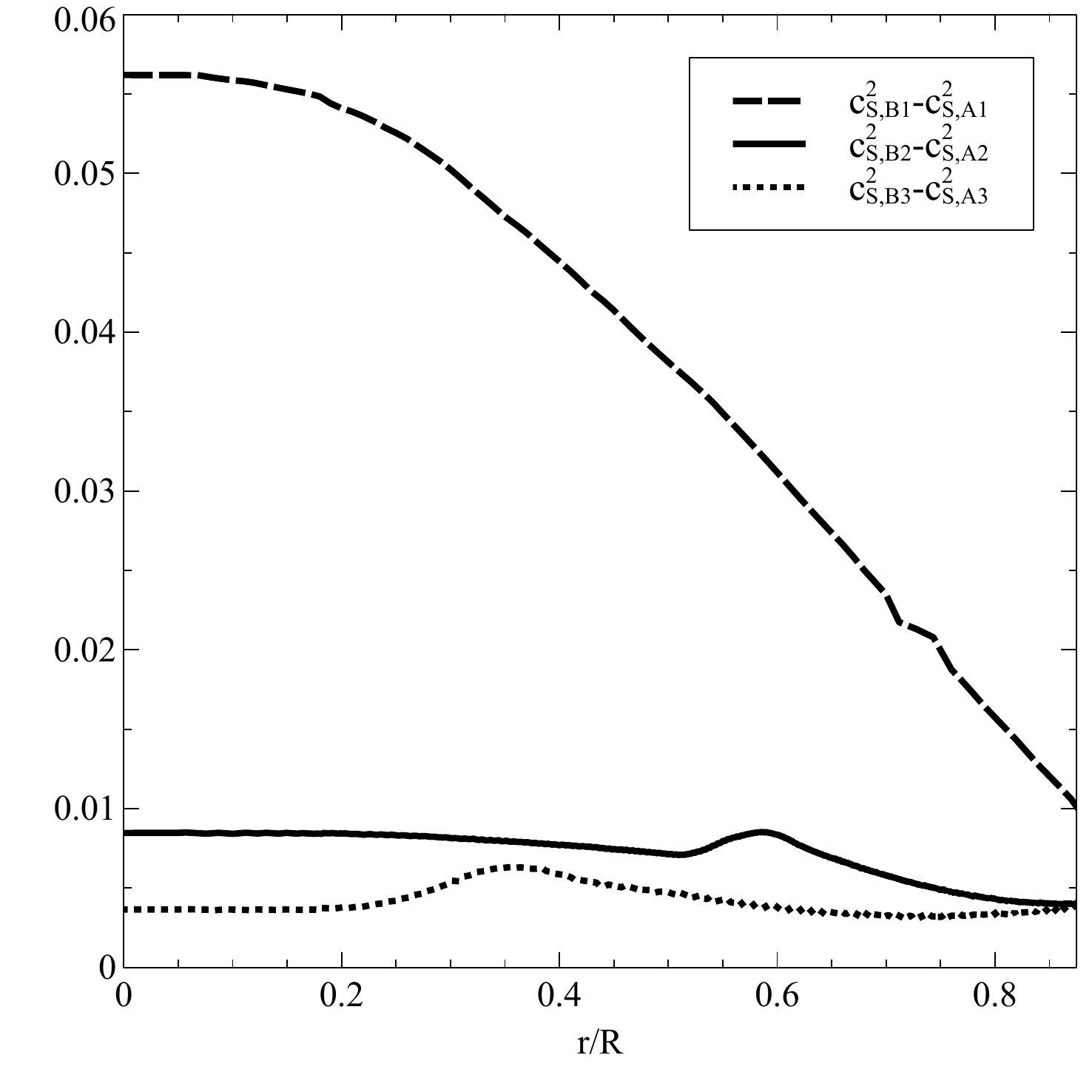} \\
\caption{{\bf Top panel:} Square of the speed of sound $c_{S}^{2}$ (for all six fiducial stars) as a function of $r/R$ (see Table \ref{tab:1set}). {\bf Lower panel:} Difference between  the $c_{S}^{2}$ for the stars with equal masses and distinctive radii that correspond to the different parameter sets (A or B) of the IST EoS as a function of $r/R$: 
$1.9 M_{\odot}$ (dashed curve), $1.5 M_{\odot}$ (solid curve) and $1.2 M_{\odot}$ (dotted curve).}                                                                                                                                                                                                                                                                                                                                                                                                                                                                                                                                                                                                                                                                                                                                                                                                                                                                                                                                                                                                                                                                                                                        
\label{fig:sound} 	
\end{figure}

\begin{figure}
\centering
\includegraphics[scale=0.45]{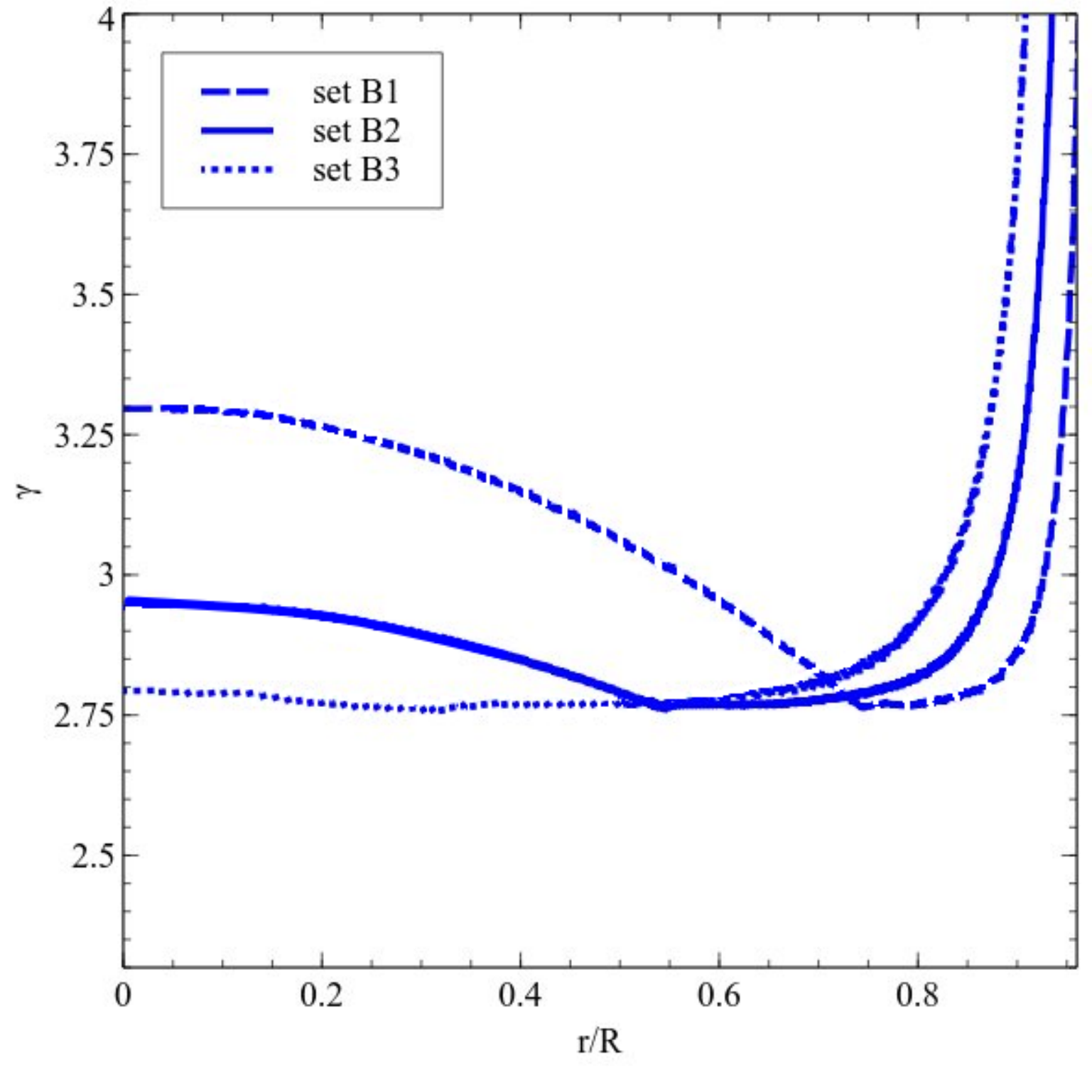} \\
\hspace{-2.0cm}\\
\includegraphics[scale=0.45]{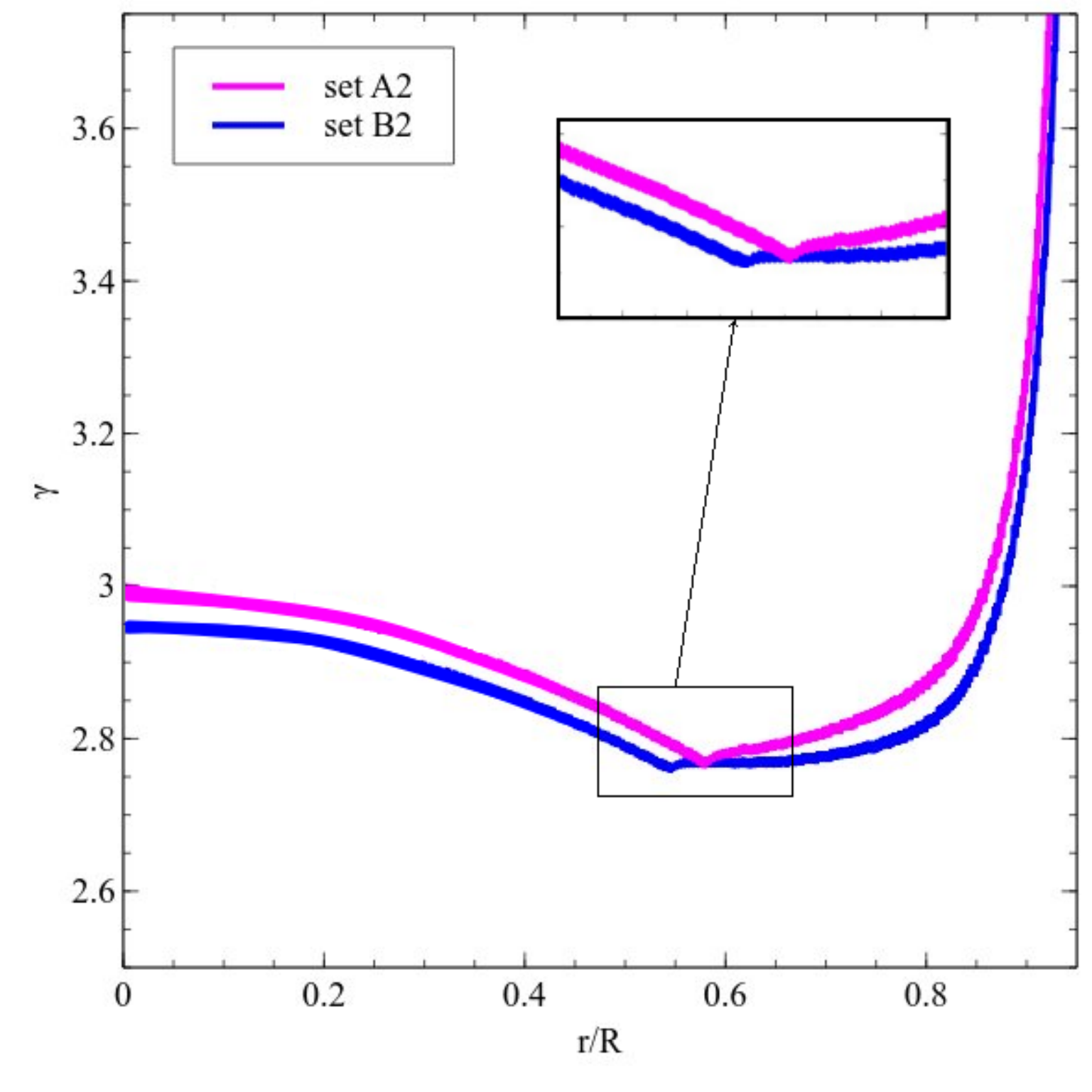} \\
\caption{ {\bf Top panel:} Adiabatic index $\gamma(r)$ as a function of $r/R$ for the stars (see Table~\ref{tab:1set}): $B_{1}$ with $M=1.9~M_{\odot}$, $B_{2}$  with $M=1.5~M_{\odot}$ and $B_{3}$ with $M=1.2~M_{\odot}$. {\bf Lower panel:} Comparison between the $\gamma$ indexes for $A_{2}$ and $B_{2}$ stars (both with a mass $M=1.5~M_{\odot}$).}                                                                                                                                                                                       
\label{fig:gamma} 
	\end{figure}
The IST contribution is a crucial term of this new EoS that accounts for the hard core repulsion effects with a very high accuracy. 
This is done by finding the correct value  of parameter $\alpha$ in order to reproduce values of the second, third and fourth virial coefficients 
of hard spheres. 
 Since the two higher virial coefficients are reproduced with only one parameter, while the second virial coefficient has the correct value for any $\alpha$ value, then we conclude that such parametrization of the hard core repulsion in the IST EoS is, indeed, physically well motivated. 
From the analysis of the virial coefficients of hard spheres it was found that $\alpha $ is approximately equal to 1.245 \citep{Sagun2017a}.
 In Ref. \citep{LS2} was confirmed that such $\alpha$ is consistent with the currently known NS properties.

The parameters of the IST EoS are determined from the fit of the different experimental observables. Parameters $C_d^2$ and $U_0$ are used in order to reproduce properties of normal nuclear matter, i.e. zero pressure and binding energy per nucleon equal to $16~MeV$ at $n=0.16~fm^{-3}$. The constants $A_{sym}$ and $B_{sym}$ are determined in order  to be in agreement with an experimental values of the nuclear asymmetry energy $J$ and its slope $L$ at nuclear saturation density. The hard core radius of nucleons $R_{nucl}$ can lie between $0.3$ and $0.5~fm$, which makes the present EoS consistent with experimental data on yields of particles produced in heavy-ion collisions (for details see \cite{violetta2}). Finally, the IST EoS has a realistic value of the nuclear incompressibility factor $K_0$, and simultaneously is consistent with the proton flow constraint \citep{Danielewicz2002} only if $\kappa=0.15 - 0.3$ \citep{Ivanytskyi2017}. The $R_{nucl}$ and $\kappa$ values were fixed by fitting this EoS to the astrophysical data. The corresponding sets of parameters are shown in Table~\ref{tab1}. Both sets of parameters provide equally realistic descriptions of the NS properties. The set B represents more softer model parameterization in comparison to the set A, that reflects in the lower values of NSs radii (see Fig. \ref{fig:massrad}), compressibilities and Love numbers. Thus, for a referent 1.4 $M_{\odot}$ NS the considered sets A and B give the Love numbers equal to 797 and 765, respectively. This result is in full agreement with the LIGO/Virgo 90\% confidence interval computed for GW170817 merger event \citep{LIGO2017}.

Having all parameters of the IST EoS fixed by requiring equal densities of electrons and protons, due to electric neutrality, as well as to the equality of the neutron chemical potential to the sum of the ones of protons and electrons (in order to ensure $\beta$-equilibrium), one can obtain a unique relation between the pressure and energy density for the NS matter.  

For simplicity, the crust was described via the polytropic EoS with $\gamma=\frac{4}{3}$. As we are not focus on the physics inside the crust of the NSs, we omitted part of the start with $r/R\gtrsim0.9$, where the transition to the crust occurs.

\section{Radial oscillations of Neutron Stars}
\label{OSC}

In the study of radial oscillations of a NS, the set of equations that describe the radial perturbations of the star matter is defined as fractional variations of the local radius $\xi=\Delta r/r$ (with $\Delta r$ being the radial displacement)
and  pressure $\eta=\Delta P/P$ (with $\Delta P$ being the perturbation of the pressure)~\citep{chanmugan, pulsating1}. Hence, 
the radial oscillations of a compact star are computed from the following system of two first-order differential equations:
\begin{eqnarray}
\xi'(r) & = & -\left(\frac{3}{r} + \frac{P'}{\zeta}\right) \xi - \frac{1}{r \gamma} \eta,\\
\eta'(r) & = & \omega^2   \left[r (1+\frac{\rho}{P}) e^{\lambda_2-\lambda_1}\right]\xi
\nonumber
 \\
&  - &  \left[ \frac{4 P'}{P} + 8 \pi \zeta r e^{\lambda_2} - \frac{r P^{'2}}{P \zeta}\right]\xi 
\nonumber
\\
& - &
 \left[ \frac{\rho P'}{P\zeta}+4 \pi \zeta r e^{\lambda_2} \right]\eta ,
\end{eqnarray}
where $e^{\lambda_1}$ and $e^{\lambda_2} = (1-2m/r)^{-1}$ are the two metric functions, $\omega$ is the frequency of the oscillation mode, $\gamma$ is the relativistic adiabatic index, that is defined by
\begin{equation}
\gamma =c_s^2 \: \left (1 + \frac{\rho}{P} \right),
\label{eq:gamma_cs}
\end{equation}
where  $c_s^2\equiv d P/d \rho$ is the adiabatic sound speed.

\smallskip

The previous system of two coupled first order differential equations is supplemented with two boundary conditions, one at the center of the star as $r \rightarrow 0$, and another at the surface $r=R$. The boundary conditions are obtained as follows: in the first equation, $\xi'(r)$ must be finite as $r \rightarrow 0$, and therefore we require that $ \eta = -3 \gamma \xi, $
must be satisfied at the center. Similarly, in the second equation, $\eta'(r)$ must be finite at the surface as $\rho,P \rightarrow 0$, and therefore we require that
\begin{equation}
\eta = \xi \left[ -4 + (1-2M/R)^{-1}  \left( -\frac{M}{R}-\frac{\omega^2 R^3}{M} \right )  \right]
\end{equation}
must be satisfied at the surface, where we recall that $M$ and $R$ are the mass and the radius of the star, respectively. 

\smallskip
As the remainder, in this article we will use the dimensionless frequency $ \sigma = {\omega}/{\omega_0} $
or  $ \nu= {\sigma \: \omega_0}/{(2 \pi)} $, where $\omega_0$ is defined by $ \omega_0 = \sqrt{{M}/{R^3}} $. 
Actually, this expression for  $ \omega_0$ gives a good estimation of frequency of  the fundamental mode. 
It is worth noticing that contrary to the previous hydrostatic equilibrium  problem (i.e.  TOV equations), which is an initial value problem, 
the problem related to the radial perturbations of a compact star,  is known as a Sturm-Liouville boundary value problem.
In this class of problems,  the frequency $\nu$ is only allowed to take particular values, the so-called eigenfrequencies $\nu_n$. Therefore, to each  specific radial oscillation mode of the star corresponds a unique $\nu_n$ (or $\sigma_n$). Accordingly, each radial mode of oscillation
is identified by its $\nu_n$ and by an associated pair of eigenfunctions $\xi_n(r)$ and  $\eta_n(r)$,
where  $\xi_n(r)$ is the displacement perturbation $\xi_n(r)$  and $\eta_n(r)$ is the pressure perturbation.


\section{Numerical results}
\label{RES}

We have considered six fiducial NSs with masses equal to $1.2 M_{\odot}$, $1.5 M_{\odot}$ and $1.9 M_{\odot}$, and radii $R \simeq (10.6-11.7)~km$ from the sets A and B (see Table~\ref{tab:1set}). 
The Fig.~\ref{fig:interior} shows typical pressure, energy density and metric functions profiles as a function of the normalised radius for the $B_{2}$ star. As presented on the top panel of  Fig.~\ref{fig:sound} the square of the speed of sound $c_{s}^{2}(r)$ decreases towards the surface of the star. For the more massive stars (sets $A_{1}$ and $B_{1}$), $c_{s}^{2}$ varies from $\sim 0.8$ 
at the center of the star to a vanishing value near the surface. To highlight the local variations of the $c_{s}^{2}$ with the star's radius for the different parameter sets of the IST EoS, the low panel of Fig.~\ref{fig:sound} shows  $c_{s,B_i}^{2}(r)-c_{s,A_i}^{2}(r)$ (for $i=1,2,3$), i.e., the difference between the square of the speed of sound for two stars with the same mass but different radius (see lower panel on Fig.~\ref{fig:sound}). It is worth noticing that all curves have a similar variation with the star's radius. The differences between the square of the speed of sound for two stars are almost flat in the core of the star, undergo a rapid increase in the layer that separates  the two regions, i.e. the inner and the outer core of the NS. The separation between these two stellar regions occur in a relatively thin transition layer located  around $0.35$,  $0.55$ or $0.75$ of the star radius that can be identified  as the protuberance in the lower panel on Fig.~\ref{fig:sound}.
To identify the physical processes responsible for such behaviour  inside the star, we computed the adiabatic index $\gamma$ as a function of the corresponding radius, that shows how the pressure varies with the baryon density. 
As first mentioned in Ref. \citep{Haensel2002}, the  analysis of $\gamma$ allows us to identify the transition layer that separates the inner and the outer core,
which, as mentioned previously, in our models occurs at $0.75$, $0.55$ and $0.35$ of the star radius for the $B_{1}$, $B_{2}$ and $B_{3}$ stars, respectively (see top panel of Fig. \ref{fig:gamma}).  As shown in the lower panel of Fig. \ref{fig:gamma},  a stiffer EoS, that  in our case corresponds 
to a IST EoS with the parameter set A, leads to a small  shift of the transition layer  to an higher radius.

\begin{table}
\begin{tabular}{l | l l l l l l}
& \multicolumn{6}{c}{{\sc Radial oscillation modes for different stars}} \\
$n$ & $A_{1}$  & $A_{2}$  & $A_{3}$  & $B_{1}$  & $B_{2}$  & $B_{3}$  \\
\hline
\hline
1  & 3.2384   & 3.2121   & 3.1758    & 3.2603  & 3.2415   & 3.2098 \\
2  & 7.2259   & 7.0942   & 6.9197    & 7.3242  & 7.1782   & 7.0102 \\
3  & 10.7881  & 10.6444  & 10.4529   & 10.9183 & 10.7763  & 10.5818 \\
4  & 14.2948  & 14.2158  & 14.0005   & 14.4571 & 14.3684  & 14.1552\\
5  & 17.8389  & 17.8093  & 17.5834   & 18.0182 & 17.9826  & 17.7572\\
6  & 21.3834  & 21.4334  & 21.2028   & 21.5757 & 21.6127  & 21.3868  \\
7  & 24.9591  & 25.0885  & 24.8530   & 25.1605 & 25.2729  & 25.0395 \\
8  & 28.5482  & 28.7557  & 28.5190   & 28.7505 & 28.9472  & 28.7121 \\
9  & 32.1512  & 32.4481  & 32.1977   & 32.3584 & 32.6368  & 32.4023\\
10 & 35.7746  & 36.1437  & 35.8918   & 35.9787 & 36.3430  & 36.1043 \\
11 & 39.3999  & 39.8581  & 39.5969   & 39.6058 & 40.0546  & 39.8154 \\
12 & 43.0433  & 43.5736  & 43.3066   & 43.2503 & 43.7822  & 43.5349 \\

\end{tabular}
\caption{Frequencies $\nu_n$ in $kHz$ for the radial modes of
 six fiducial stars considered here (see Table~\ref{tab:1set}). $n$ is the  order of the radial mode.}
\label{tab:2set}
\centering
\end{table}

\begin{figure}
\centering
\includegraphics[scale=0.45]{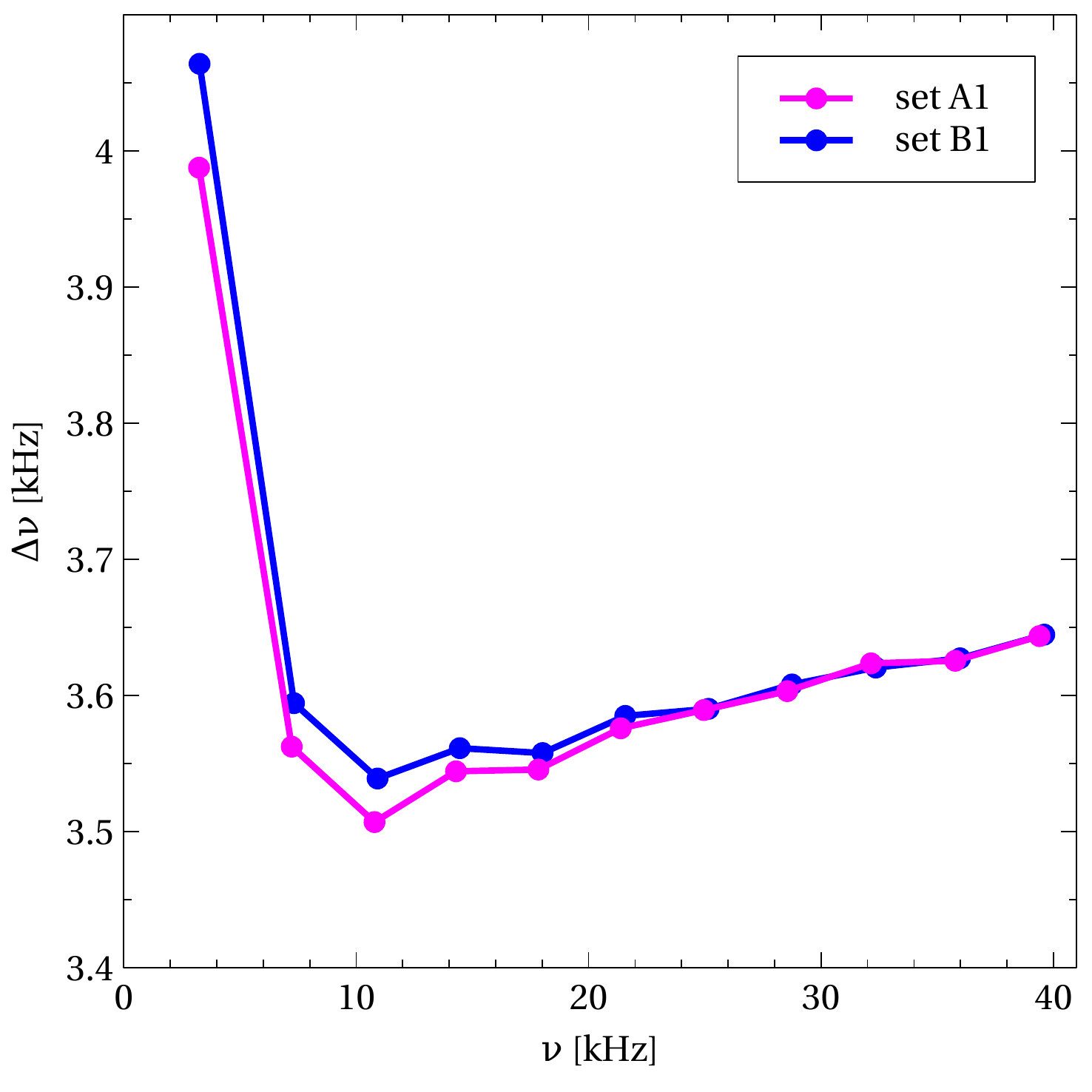} \\
\includegraphics[scale=0.45]{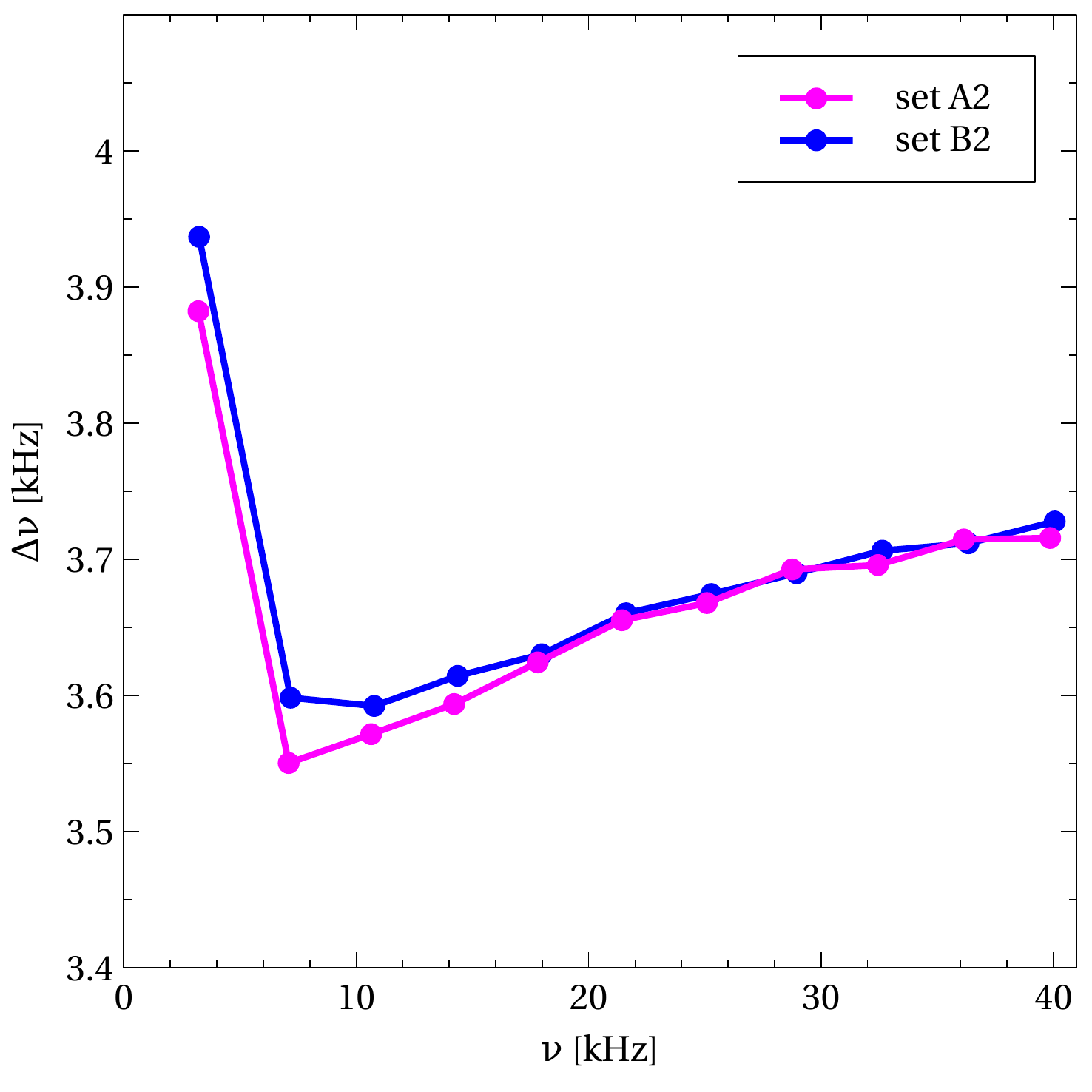} \\
\includegraphics[scale=0.45]{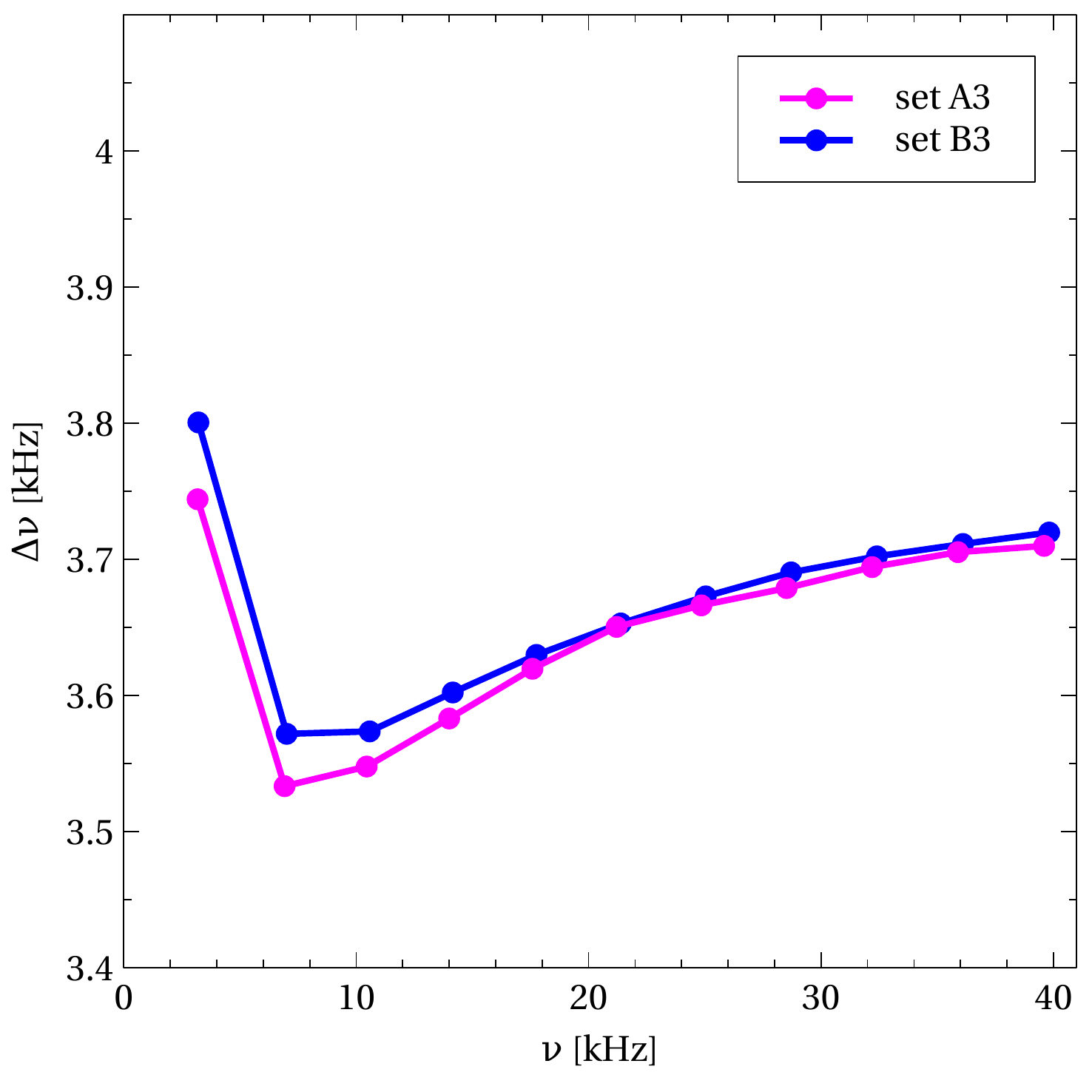} 
\caption{
Large frequency difference vs frequency (both in kHz) for two stars of the same mass from sets $A$ (in magenta) and $B$ (in blue). {\bf Top panel:} Comparison between stars $A_{1}, B_{1}$ with mass $M=1.9~M_{\odot}$. {\bf Middle panel:} Comparison between stars $A_{2}, B_{2}$ with mass $M=1.5~M_{\odot}$. {\bf Lower panel:} Comparison between stars $A_{3}, B_{3}$ with mass $M=1.2~M_{\odot}$.}                                                                                                                                                                                                                                                                                                                                                                                                                                                                                                                                                                                                                                                                                                                                                                                                                                                                                                                                                                                                                                                                                                                        
\label{fig:spectra} 	
\end{figure}

\smallskip

Table~\ref{tab:2set} shows the frequencies of the first 12 radial modes for the fiducial stars of Table~\ref{tab:1set}.
One of the quantities widely used in asteroseismology to learn about star properties is a difference between consecutive modes, i.e. $\Delta\nu_{n}=\nu_{n+1}-\nu_n$, the so-called large separation~\citep{book,lopes}. In Fig. \ref{fig:spectra}  we show the comparison of 
the $\Delta\nu_{n}$ functions for every pair of stars with the same mass. 
In general, 
we found that the decrease of the central baryon density, and, therefore,  of the star's mass leads to a decrease of the large separation $\Delta\nu_{n}$. Interestingly, it was also found that $\Delta\nu_{n}$ varies with $\nu_{n}$ 
which is a first evidence that the microphysics (or the EoS) of the interior of the NS is imprinted in the large separation, a characteristic well-known and  also found in main sequence stars, 
for instance the Sun. 	
For both types of stars,  $\Delta\nu_{n}$ starts to be a constant 
proportional to $\sqrt{M/R^3}$ that is independent of $\nu_n$, 
on top of which a discontinuity or  glitch  on the star's structure 
will imprint a small $\nu_n-$ oscillation on 
$\Delta\nu_{n}$, like the one observed in Fig.~\ref{fig:spectra}.		
In the case of the NS, this $\nu_n-$ oscillation has an amplitude proportional to  the magnitude of the discontinuity, which results from the rapid variation
of the sound speed $c_s$ or the relativistic adiabatic index  $\gamma$
(see equation \ref{eq:gamma_cs}) on the transition layer that separates the inner and the outer core of the NS,
as clearly shown in  Fig. ~\ref{fig:gamma}. 
The period of the $\nu_n-$ oscillation relates to the location of the discontinuity beneath the star's surface~\cite{LopesTC}. 
A detailed account for the impact of discontinuities or glitches on radial and non-radial acoustic oscillations is described on 
 \cite[e.g.,][]{lopesgough,BritoLopes2014}.

\begin{figure}
\centering
\includegraphics[scale=0.5]{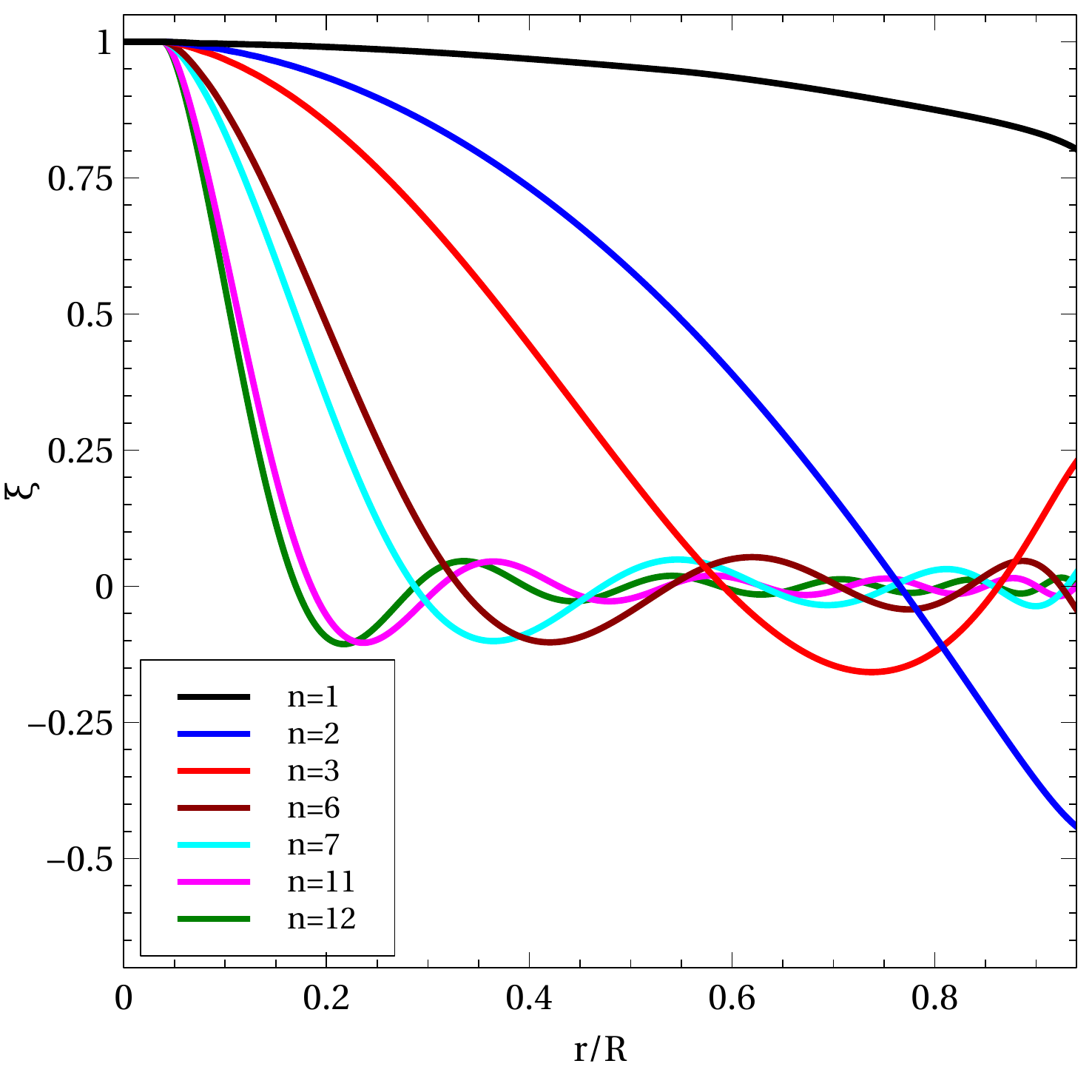} \\
\includegraphics[scale=0.5]{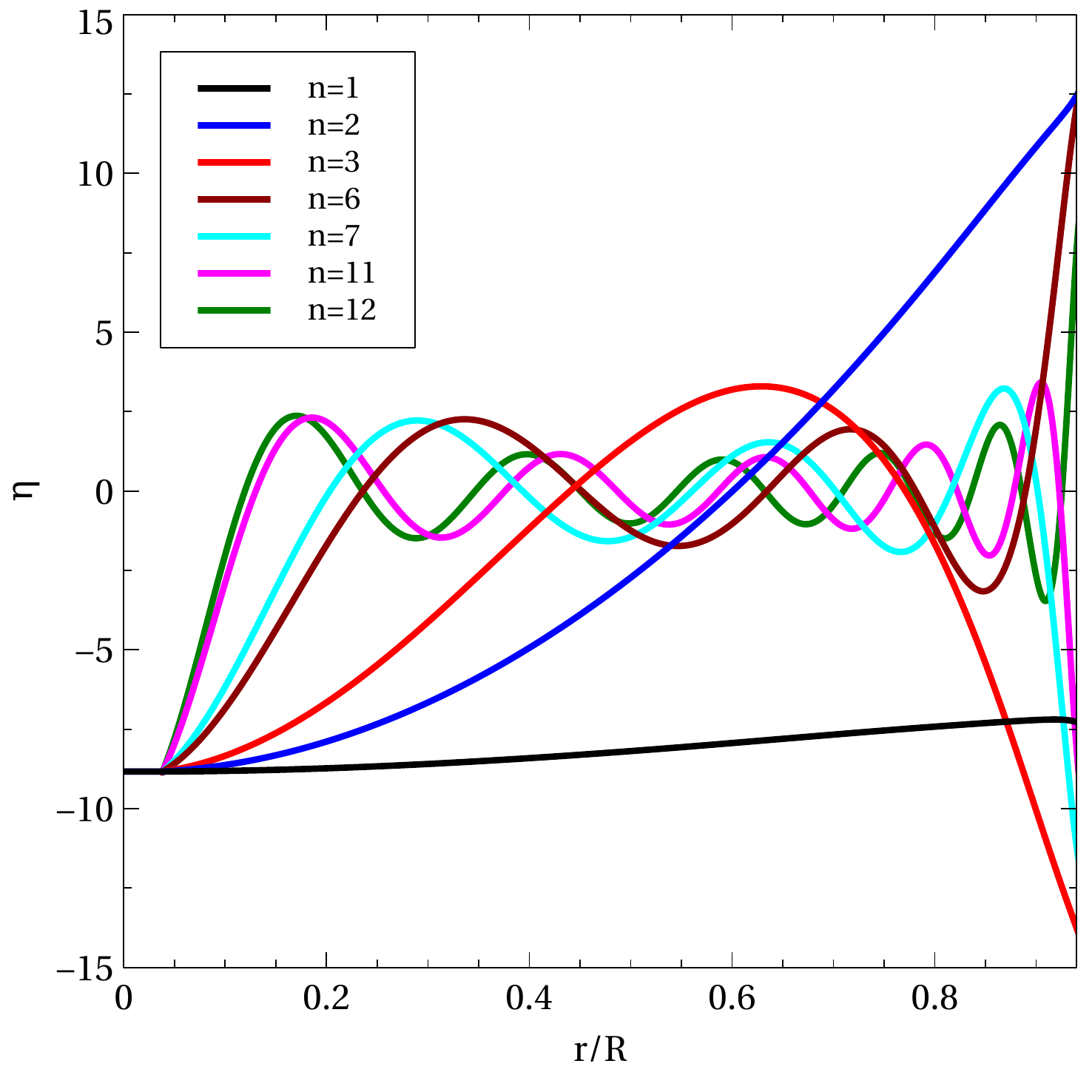} 
\caption{
{\bf Top panel:} First eigenfunction $\xi_n$ vs dimensionless radius coordinate $r/R$ for low ($n=1,2,3$), intermediate ($n=6,7$) and highly-excited modes ($n=11,12$). {\bf Lower panel:} Same as before, but for the second eigenfunction $\eta_n$. Both plots correspond to $B_{2}$ star.}
\label{fig:oscillations} 	
\end{figure}

On Fig.~\ref{fig:oscillations} it  is shown the $\xi_n$  and  $\eta_n$ eigenfunctions  calculated for the $B_{2}$ star (with $M=1.5 M_{\odot}$ and $R=11.31~km$). These eigenfunctions are computed  for the low order radial modes, $n=1,2,3$, shown in black, blue and red, respectively, intermediate modes, $n=6,7$, shown in dark red and cyan, and finally highly excited modes, $n=11,12$, shown in magenta and green. 
According to a Sturm-Liouville boundary value problem the number of zeros of the eigenfunctions corresponds to the overtone number $n$, namely the first excited mode, corresponding to $n=2$, has only one zero, the second excited mode, corresponding to $n=3$, has two zeros, while the fundamental mode, corresponding to $n=1$, does not have any zeros at all. The fundamental mode is also known as the f-mode, while the rest of the modes with $n=2,3,...$ are the so-called p-modes (pressure modes or acoustic modes)~\citep{book}. 

The amplitude of $\xi_n(r)$ for each mode $n$ is larger closer to the center (but not at the center) and much smaller near the surface. Alternatively, the amplitude of $\eta_n(r)$  is larger closer to the center and near the surface of the star. Hence, it results that $\xi_{n+1}(r)-\xi_n(r)$ and $\eta_{n+1}(r)-\eta_n(r)$ are more sensitive to the core of the star.  Notice, that although  $\eta_n(r)$ of consecutive $n$ have large amplitudes near the surface with opposite signal (opposition of phase) its contribution for $\eta_{n+1}(r)-\eta_n(r)$ cancels out.

\begin{figure}
\centering
\includegraphics[scale=0.5]{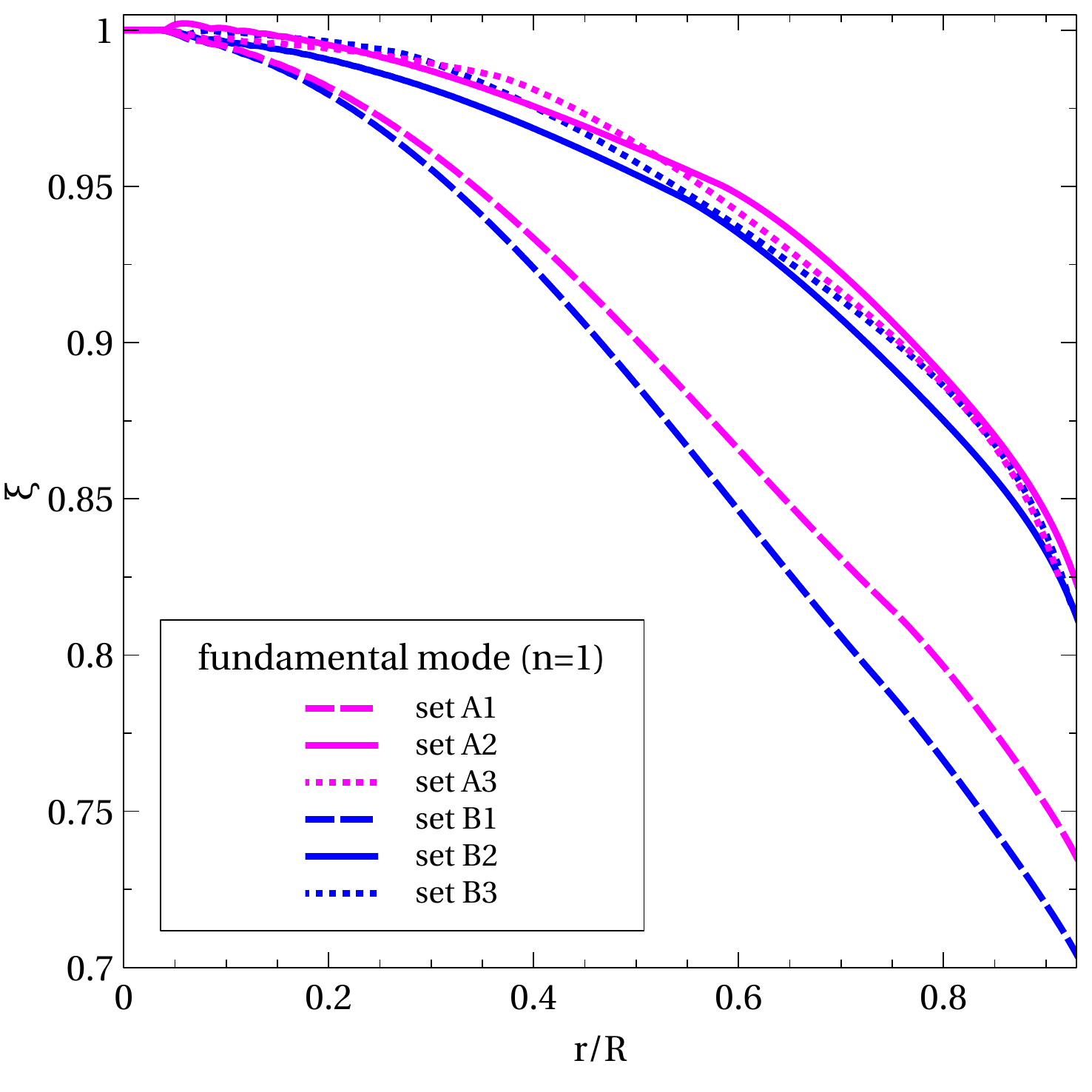} \\
\caption{
Comparison between the first eigenfunctions $\xi_n$ for the fundamental mode (n=1) for all six considered stars.}
\label{fig:ksi} 	
\end{figure}

As one can see on Fig. \ref{fig:ksi}, the comparison of the fundamental modes for all six considered stars reveals an imprint of the stars structure on the behaviour of the first eigenfunction with star radius. Humps on the eigenfunction profiles are shifted to higher radius with an increase of the star mass. As a result, for $B_{1}$, $B_{2}$, $B_{3}$ stars it happens at about $0.75$, $0.55$ and $0.35$ of star radius, respectively. The onset of anomaly
is  become shifted to higher $r/R$ values for a stiffer IST EoS (set A, see details on Fig. \ref{fig:ksi}). Smooth behaviour of the first eigenfunction for all higher oscillation modes can be explained by its negligible effect for modes with a bigger n.

It is possible to conclude that changes of the thermodynamic properties of the matter inside the NSs, i.e., variations in the IST EoS  leaves an imprint on the $\xi_n(r)$ eigenfunction for the f-mode. Such changes of star properties correspond to the different layers of the star. Thus, we conclude that the found irregularities are associated with the transition between the inner and the outer core of star. Moreover, we found that for a more massive star (e.g., $B_{1}$ on Fig. \ref{fig:ksi} and top panel of Fig. \ref{fig:gamma}), the transition layer occurs  closer to the  star's surface, where for a low mass NS
it occurs more closer to the center, i.e. around $0.35$ of the radius of the star. This result fundaments our point that if radial oscillations of the NSs are discovered, 
it will be possible to use frequencies of radial modes to learn about the thermodynamic properties of the matter inside NS.

As you can see in the Table \ref{tab:2set}, the oscillation frequencies grow with an increase of the central baryon density (see Table \ref{tab:1set}). As was discussed in Refs. \cite{pulsating8, Zdunik1999} the oscillation frequency of the first mode start to decrease while approaching the maximal mass (or central density) for a given EoS and cross a zero value exactly at the highest point of a M-R curve.

\section{Detectability and excitation mechanisms of the NS oscillations}
\label{Detect}

Study of the physical mechanisms leading to excitation of the NS oscillations is a very challenging problem due to the  interplay between thermodynamical properties of the NS matter, mass of the star, magnetic field, spin, etc. Recent studies suggest that the oscillations of a hypermassive NS (HMNS), formed as a result of a NSs merger, will create a modulation of a short gamma ray burst (SGRB) signal, which is possible to detect \cite{Chirenti2019}.

The fundamental f-modes can also be excited by tidal effects in close eccentric systems \cite{Chirenti2017}, as well as due to a resonant excitation in binaries \cite{Hinderer}. Between the other probable mechanisms to excite such oscillations there are accretion in Low-Mass X-ray Binaries (LMXBs) \cite{LMXRB}, magnetic reconfiguration, during the supernova explosion, etc. \cite{Franco2000,Tsang2012}.

Unfortunately, low sensitivity of the ongoing gravitational wave detectors at kHz frequency range does not allow the detection of the NS oscillations. However, the third-generation of ground-based gravitational wave detectors, e.g. the Einstein Telescope and the Cosmic Explorer \cite{GWdetectors, Chirenti2017}, are expected to have a sensitivity much higher than an order of magnitude in comparison to the Advanced LIGO. Such detections could provide with simultaneous measurements of NS masses, tidal Love numbers, frequency, damping time, amplitude of the modes, and, therefore, moments of inertia, which will give an observational opportunity to test the I-Love-Q relation \cite{Chirenti2017,ILoveQ}.


\section{Conclusions}
\label{Concl}

We have studied radial oscillations of NSs within an elaborate IST EoS, which is in a good agreement with the normal nuclear matter properties, provides a high quality description of the proton flow constraint, hadron multiplicities created during the nuclear-nuclear collision experiments and equally is consistent with astrophysical data coming from NS observations and the GW170817 NS-NS merger. We have considered six fiducial stars with masses $M=1.9~ M_{\odot}$, $M=1.5~M_{\odot}$ and $M=1.2~M_{\odot}$, and radii $R\simeq(10.6-11.7)~km$, from two different sets (A and B) of model parameters. For all six considered stars we have computed 12 lowest radial oscillation modes, the large frequency separations and the corresponding eigenfunctions. It was shown that softer IST EoS (with parameter set B)  in comparison to the stiffer IST EoS (with parameter set A)
 gives larger frequencies of oscillations  for all considered stars.  Accordingly, the large frequency separation
 $\Delta\nu_{n}$ for both sets has the same behaviour. Similarly, for the same model set the calculated frequencies also grow with an increase of the central baryon density.
 
Moreover,  we found an evidence of how the changes in thermodynamic properties of the NS matter leave an imprint on the $\xi_n(r)$ eigenfunctions calculated for a fundamental mode (n=1). Analysis of the adiabatic index $\gamma (r)$ and speed of sound $c_s(r)$ shows that clear and well defined  changes occur inside the NS that we found to be associated with a transition layer between the inner and outer core of the star. For example, for $B_{1}$, $B_{2}$, $B_{3}$ stars, such transition occurs at the locations of $0.75$, $0.55$ and $0.35$ of the star radius, respectively. The $\xi_n(r)$ eigenfunction calculated for the fundamental mode presents changes of behaviour exactly at the same values of star's radius where the adiabatic index $\gamma (r)$ and the
difference between  the speed of sound squared $c_{S}^{2}$ for the stars with equal masses and distinctive radii  show such peculiar behaviour.

The results found in this work exhibit an imprint of the thermodynamic properties of matter and internal structure of NS on the radial oscillation modes, and, possibly, even non-radial modes, as a similar global behaviour is expected.
Furthermore, coupling between the radial and non-radial oscillations that leads to the enhanced gravitational emission makes it possible to detect such oscillations during the NS-NS merger. 
We predict a similar analysis for a model with a more realistic description of the NS crust and inclusion of the quark-gluon core, could reveal other more prominent irregularities in the oscillation frequencies and the eigenfunctions calculated for the different oscillation modes. 

Finally, we discuss the main known mechanisms to excite oscillation modes and the probability of their detection with a third-generation of ground-based gravitational wave detectors, such as the Einstein Telescope and the Cosmic Explorer.

\section*{Acknowledgements}

We thankful for the fruitful discussions and suggestions by A. Brito, C. Chirenti, O. Ivanytskyi and J. L. Zdunik. V.S. acknowledges for the partial financial support from the Funda\c c\~ao para a Ci\^encia e Tecnologia (FCT), Portugal, for the support through the grant No. UID/FIS/04564/2019 and the Program of Fundamental Research in High Energy and Nuclear Physics launched by the Section of Nuclear Physics of the National Academy of Sciences of Ukraine. G.P and I.L. thank the Funda\c c\~ao para a Ci\^encia e Tecnologia (FCT), Portugal, for the financial support to the Center for Astrophysics and Gravitation-CENTRA,  Instituto Superior T\'ecnico, Universidade de Lisboa, through the grant No. UID/FIS/00099/2013. This research was supported by the Munich Institute for Astro- and Particle Physics (MIAPP) which is funded by the Deutsche Forschungsgemeinschaft (DFG, German Research Foundation) under Germany`s Excellence Strategy \--- EXC-2094 \--- 390783311. 




\begin{thebibliography}{99}

\bibitem{LIGO2017}
B.~P. Abbott {\it et al.} (LIGO Scientific Collaboration and Virgo Collaboration), Phys. Rev. Lett., {\bf 121}, 161101 (2018).

\bibitem{LMXRB}
N. Andersson, D.~I. Jones, K.~D. Kokkotas, N. Stergioulas, ApJ, {\bf 534}, L75 (2000).

\bibitem{pulsating8}
A. Brillante, I.~N. Mishustin, EPL, {\bf 105}, 3, 39001 (2014).


\bibitem{2018NuPhA.970..133B}
K.~A. Bugaev, V.~V. Sagun, A.~I. Ivanytskyi, I.~P. Yakimenko {\it et al.}, Nucl.\ Phys.\ A, {\bf 970}, 133 (2018).

\bibitem{Bugaev1989}
K.~A. Bugaev, M.~I. Gorenstein, Z.\ Phys.\ C\, {\bf 43}, 261 (1989).


\bibitem{BritoLopes2014}
A. Brito, I. Lopes, ApJ, {\bf 782}, article id. 16 id. 16 (2014).


\bibitem{chanmugan}
G. Chanmugan, ApJ, {\bf 217}, 799 (1977).

\bibitem{Chirenti2019}
C. Chirenti, M.~C. Miller, T. Strohmayer, J. Camp, arXiv:1906.09647 [astro-ph.HE] (2019).

\bibitem{Chirenti2018}
C. Chirenti, M. Jasiulek, Mon.\ Not.\ R.\ Astron.\ Soc.\, {\bf 476}, 354 (2018).

\bibitem{Chirenti2017}
C. Chirenti, R. Gold, M.~C. Miller, ApJ, {\bf 837}, 67 (2017).

\bibitem{Danielewicz2002}
P. Danielewicz, R. Lacey, W.~G. Lynch, Science, {\bf 198}, 1592 (2002).

\bibitem{GR}
A. Einstein, Annalen Phys., {\bf 49}, 769 (1916).

\bibitem{Franco2000}
L.~M. Franco, B. Link, R.~I. Epstein, ApJ, {\bf 543}, 987 (2000).

\bibitem{Zdunik1999}
D. Gondek, J.~L. Zdunik, Astron.\ Astrophys.\, {\bf 344}, 117 (1999).

\bibitem{Haensel2002}
P. Haensel, K.~P. Levenfish, D.~G. Yakovlev, Astron.\ Astrophys.\, {\bf 394}, 213 (2002).

\bibitem{HaenselPotekhin2004}
P. Haensel, A.~Y. Potekhin, A\&A {\bf 428}, 191 (2004).

\bibitem{Hinderer}
T. Hinderer, A. Taracchini, F. Foucart, A. Buonanno, Phys.\ Rev.\ Lett.\, {\bf 116} 181101 (2016).


\bibitem{Zand2019}
J.~J.~M. in't Zand {\it et al.}, Sci.\ China-Phys.\ Mech.\ Astron.\, {\bf 62}, 029506 (2019).

\bibitem{Ivanytskyi2017}
A.~I. Ivanytskyi, K.~A. Bugaev, V.~V. Sagun, L.~V. Bravina {\it et al.}, Phys.\ Rev.\ C\, {\bf 97}, 064905 (2018).

\bibitem{pulsating2}
K.~D. Kokkotas, J. Ruoff, Astron.\ Astrophys.\, {\bf 366}, 565 (2001).

\bibitem{strangestars1}
D. Leahy, R. Ouyed, Mon.\ Not.\ R.\ Astron.\ Soc.\, {\bf 387}, 1193 (2008).


\bibitem{LopesTC}
I.~P. Lopes, S. Turck-Chi\'eze, Astron.\ Astrophys.\, {\bf 290}, 845 (1994).
 
\bibitem{lopes}
I.~P. Lopes, Astron.\ Astrophys.\, {\bf 373}, 916 (2001).
 
 \bibitem{lopesgough}
 I.~P. Lopes, D. Gough
 Monthly Notices of the Royal Astronomical Society,{\bf 322},  473 (2001).
 
 
\bibitem{pulsating3}
G. Miniutti, J.~A. Pons, E. Berti, L. Gualtieri {\it et al.}, Mon.\ Not.\ R.\ Astron.\ Soc.\, {\bf 338}, 389 (2003).

\bibitem{TOV2}
J.~R. Oppenheimer, G.~M. Volkoff, Phys.\ Rev.\, {\bf 55}, 374 (1939).

\bibitem{strangestars2}
R. Ouyed, D. Leahy, P. Jaikumar, arXiv:0911.5424 [astro-ph.HE] (2009).
  
\bibitem{Ozel2006}
F. {\"O}zel, P. Freire,  Astron.\ Astrophys.\, {\bf 54}, 401 (2016).
  

\bibitem{pulsating10}
G. Panotopoulos, I. Lopes, Phys.\ Rev.\ D, {\bf 96}, 8, 083013 (2017).

\bibitem{pulsating5}
A. Passamonti, M. Bruni, L. Gualtieri, A. Nagar {\it et al.}, Phys.\ Rev.\ D, {\bf 73}, 084010 (2006).

\bibitem{pulsating4}
A. Passamonti, M. Bruni, L. Gualtieri, C.~F. Sopuerta, Phys.\ Rev.\ D\, {\bf 71}, 024022 (2005).

\bibitem{GWdetectors}
T. Regimbau, M. Evans, N. Christensen, E. Katsavounidis {\it et al.}, Phys.\ Rev.\ Lett.\, {\bf 118} 151105 (2017).

\bibitem{Rischke1988}
D.~H. Rischke, B.~L. Friman, H. St{\"o}cker, W. Greiner,  J.\ Phys.\ G\, {\bf 14}, 191 (1988).


\bibitem{LS2}
V.~V. Sagun, I. Lopes, A.~I. Ivanytskyi, Astrophys.\ J.\, {\bf 871}, 157 (2019).
  
\bibitem{violetta2}
V.~V. Sagun, K.~A. Bugaev, A.~I. Ivanytskyi, I.~P. Yakimenko {\it et al.}, Eur.\ Phys.\ J.\ A, {\bf 54}, 6, 100 (2018).

\bibitem{Sagun2017a}
V.~V. Sagun, K.~A. Bugaiev, A.~I. Ivanytskyi, D.~R. Oliinychenko {\it et al.}, Eur.\ Phys.\ J.\ Web\ Conf.\, {\bf 137}, 09007 (2017).

\bibitem{LS1}
V.~V. Sagun, I. Lopes, Astrophys.\ J.\, {\bf 850}, 1, 75 (2017).
  
\bibitem{violetta1}
V.~V. Sagun, A.~I. Ivanytskyi, K.~A. Bugaev, I.~N. Mishustin, Nucl.\ Phys.\ A, {\bf 924}, 24 (2014).

\bibitem{pulsating6}
G.~J. Savonije, Astron.\ Astrophys.\, {\bf 469}, 1057 (2007).

\bibitem{SBH}
K. Schwarzschild, Sitzungsber.\ Preuss.\ Akad.\ Wiss.\ Berlin (Math.\ Phys.)\, 189 (1916).

\bibitem{textbook}
S.~L. Shapiro, S.~A. Teukolsky, New York, USA: Wiley 645 (1983).

\bibitem{Steiner2013}
A.~W. Steiner, J.~M. Lattimer, E.~F. Brown,  Astrophys.\ J.\, {\bf 765}, 5 (2013).

\bibitem{Steiner2010}
A.~W. Steiner, J.~M. Lattimer, E.~F. Brown, Astrophys.\ J.\, {\bf 722}, 33 (2010).

\bibitem{TOV1}
R.~C. Tolman, Phys.\ Rev.\, {\bf 55}, 364 (1939).

\bibitem{Tsang2012}
D. Tsang, J.~S. Read, T. Hinderer, A.~L. Piro {\it et al.}, Phys.\ Rev.\ Lett.\, {\bf 108}, 011102 (2012).

\bibitem{book}
W. Unno, Y. Osaki, A. Hiroyasu, H. Saio {\it et al.}, Tokyo: University of Tokyo Press, 2nd ed. (1989).

\bibitem{pulsating9}
C. V{\'a}squez Flores,  Z.~B. Hall ~II, P. Jaikumar, Phys.\ Rev.\ C\, {\bf 96}, 6, 065803 (2017).

\bibitem{pulsating7}
C. V{\'a}squez Flores, G. Lugones, Phys.\ Rev.\ D\, {\bf 82}, 063006 (2010).

\bibitem{pulsating1}
H.~M. V{\"a}th, G. Chanmugan, Astron.\ Astrophys.\, {\bf 260}, 250 (1992).

\bibitem{ILoveQ}
K. Yagi, N. Yunes, Phys.\ Rev.\ D\, {\bf 88} 023009 (2013).


\bibitem{Symslope}
Z. Zhang,  L.-W. Chen, Phys.\ Lett.\ B \, {\bf 726}, 234 (2013).


\end{thebibliography}
\end{document}